\documentclass[iop,twocolappendix]{emulateapj}

\usepackage{verbatim,float}

\newcommand{\zwindow}{$2.5<z<3.0$}
\newcommand{\zmed}{$z \approx 2.6$}
\newcommand{\mmed}{$M \approx 10^{11.28}~M_{\odot}$}
\newcommand{\mcut}{11.25}
\newcommand{\msun}{$M_{\odot}$}
\newcommand{\masslimit}{$M \geq 10^{\mcut}~M_{\odot}$}
\newcommand{\vdw}{vdW14}
\newcommand{\nstar}{21}
\newcommand{\betaz}{-0.90 \pm 0.12}
\newcommand{\betaH}{-0.85 \pm 0.12}
\newcommand{\plotdir}{}

\begin{document}

\title{A comparison of the most massive quiescent galaxies from $\lowercase{z} \sim 3$ to the present: slow evolution in size, and spheroid-dominated\altaffilmark{1}}

\author{Shannon G. Patel\altaffilmark{2}, Yu Xuan Hong\altaffilmark{2,3}, Ryan F. Quadri\altaffilmark{4}, Bradford P. Holden\altaffilmark{5}, Rik J. Williams\altaffilmark{2,6}}
\affil{\altaffilmark{2}Carnegie Observatories, 813 Santa Barbara Street, Pasadena, CA 91101, USA, e-mail: patel@obs.carnegiescience.edu}
\affil{\altaffilmark{3}Pomona College, Claremont, CA 91711, USA}
\affil{\altaffilmark{4}Department of Physics \& Astronomy, Texas A\&M University, College Station, TX 77843, USA}
\affil{\altaffilmark{5}UCO/Lick Observatory, University of California, Santa Cruz, CA 95064, USA}
\affil{\altaffilmark{6}Uber Technologies Inc., 1455 Market St, 4th Floor, SF CA 94103, USA}

\begin{abstract}
We use {\em Hubble Space Telescope} imaging to study the structural properties of ten of the most massive (\masslimit) quiescent galaxies (QGs) in the UKIDSS UDS at \zwindow.  The low spatial density of these galaxies required targeted WFC3 $H_{160}$ imaging, as such systems are rare in existing surveys like CANDELS.  We fit Sersic models to the 2D light profiles and find that the median half-light radius is $R_e \sim 3$~kpc, a factor of $\sim 3$ smaller than QGs with similar masses at $z \sim 0$.  Complementing our sample with similarly massive QGs at lower redshifts, we find that the median size evolves as $R_e \propto H(z)^{\betaH}$ (or alternatively, $R_e \propto (1+z)^{\betaz}$).  This rate of evolution is slower than that for lower mass QGs.  When compared to low redshift QGs, the axis ratio distribution for our high redshift massive QG sample is most consistent with those in which spheroids are dominant.  These observations point to earlier size growth among massive QGs that also resulted in spheroidal systems.  Finally, we measured residual-corrected surface brightness profiles for our sample.  These show that the Sersic parameterization is generally representative out to several effective radii and does not miss excess low surface brightness light.  The sizes inferred from the light profiles therefore confirm the compactness of these most massive high redshift QGs.

\end{abstract}

\altaffiltext{1}{Based on observations made with the NASA/ESA Hubble Space Telescope, obtained at the Space Telescope Science Institute, which is operated by the Association of Universities for Research in Astronomy, Inc., under NASA contract NAS 5-26555. These observations are associated with program \#13002.}


\section{Introduction}

The most massive galaxies in the nearby universe generally lack star formation and have early-type morphologies \citep[e.g.,][]{brinchmann2004,kelvin2014b}.  In recent years, such quiescent galaxies (QGs) have been observed to $z \sim 1$ and beyond where measurements of their half-light radii revealed systems that were much more compact \citep[e.g.,][]{daddi2005,trujillo2006b,zirm2007,toft2007,newman2012}.  Their compact nature appears robust as measurements of their high stellar masses have been corroborated through spectroscopic studies of their kinematics \citep[e.g.,][]{vanderwel2008c,cappellari2009b,newman2010,vandesande2011,vandesande2013}.  In addition, deep WFC3 observations have not revealed any missing extended low surface brightness light from these QGs that could have been masked by noise in shallower imaging \citep{szomoru2010,szomoru2011,szomoru2012,szomoru2013}.  

The interpretation of the observed size evolution of QGs is the source of considerable discussion and debate.  The most commonly cited explanation is that they grow through dissipationless mergers with smaller galaxies that are tidally disrupted and deposit their stars in an envelope around the larger primary \citep{naab2009b,bezanson2009,hilz2013}.  Coincident with this picture, progenitor matching studies have shown that stars in massive galaxies assemble inside-out, leading to an increase in the half-light radius to low redshift \citep{vandokkum2010,patel2013,morishita2015,davari2016}.  Recent cosmological simulations also find that, in the majority of cases, high redshift compact QGs end up as the cores of low redshift ellipticals \citep{wellons2016}.  However, the ongoing transfer of newly quenched galaxies onto the red sequence may also be responsible for some portion of the size growth, as these galaxies tend to have larger sizes toward low redshift \citep[e.g.,][but for an opposing view see \citet{whitaker2012}]{carollo2013,belli2015}.

Another intriguing aspect of massive high redshift QGs is their shape.  Their axis-ratio distribution indicates the presence of disks \citep{vanderwel2011,chang2013b}.  This stands in stark contrast to their counterparts today, which are spheroidal \citep{vanderwel2009b,holden2012}.  If dry minor mergers are indeed a dominant mechanism for size growth, a convenient byproduct of such a process is the destruction of disks and morphological transformation to spheroids \citep{bournaud2007c}.

Despite the wealth of information in the literature pertaining to high redshift QGs, samples are severely limited for the most massive QGs.  For example,  \citet[][hereafter referred to as \vdw]{vanderwel2014} combined all five CANDELS/3D-HST fields in their extensive size-mass study but only found three QGs above \masslimit\ at \zwindow.  Larger samples at high redshift are therefore needed to understand the evolution of this high mass population relative to those at lower masses.

Owing to their low spatial density, massive QGs at high redshift require targeted HST observations.  In this paper, we present new WFC3 $H_{160}$ imaging for ten QGs with stellar mass \masslimit\ at \zwindow\ in the UKIRT Infrared Deep Sky Survey (UKIDSS) Ultra-Deep Survey (UDS) \citep{lawrence2007}.  In Section~\ref{sec_data} we present the data sets as well as the relevant measurements employed in this work.  In Section~\ref{sec_results} we discuss the half-light radii and axis ratios for our sample of high redshift massive QGs.  We compare these properties to other QGs spanning a range of masses and redshifts.  We also measure surface brightness profiles and compare them to a local QG sample from SDSS.  Finally, we discuss our findings in Section~\ref{sec_discussion} and summarize our conclusions in Section~\ref{sec_summary}.

We assume a cosmology with $H_0=70$~km~s$^{-1}$~Mpc$^{-1}$, $\Omega_M=0.3$ and $\Omega_{\Lambda}=0.7$.  Stellar masses are based on a \citet{chabrier2003} IMF.  All magnitudes are given in the AB system.

\section{Data \& Analysis} \label{sec_data}

\subsection{Massive quiescent galaxies selected in the UDS field} \label{sec_selection}

We selected targets from the UKIDSS UDS catalog (DR8) presented by \citet{williams2009,williams2010} and \citet{quadri2012}.  We refer the reader to those works for details on the data reduction, photometry, and SED fitting, and briefly summarize the pertinent points here.  Objects were selected from the portion of the field containing full optical and near-IR imaging which spans $\sim 0.65$~deg$^2$.  Photometric redshifts were computed with EAZY \citep{brammer2008} and stellar masses were measured using FAST \citep{kriek2009} with exponentially declining, solar-metallicity, star formation histories in combination with a \citet{calzetti2000} reddening law.

\begin{figure}
\epsscale{1.2}
\plotone{\plotdir 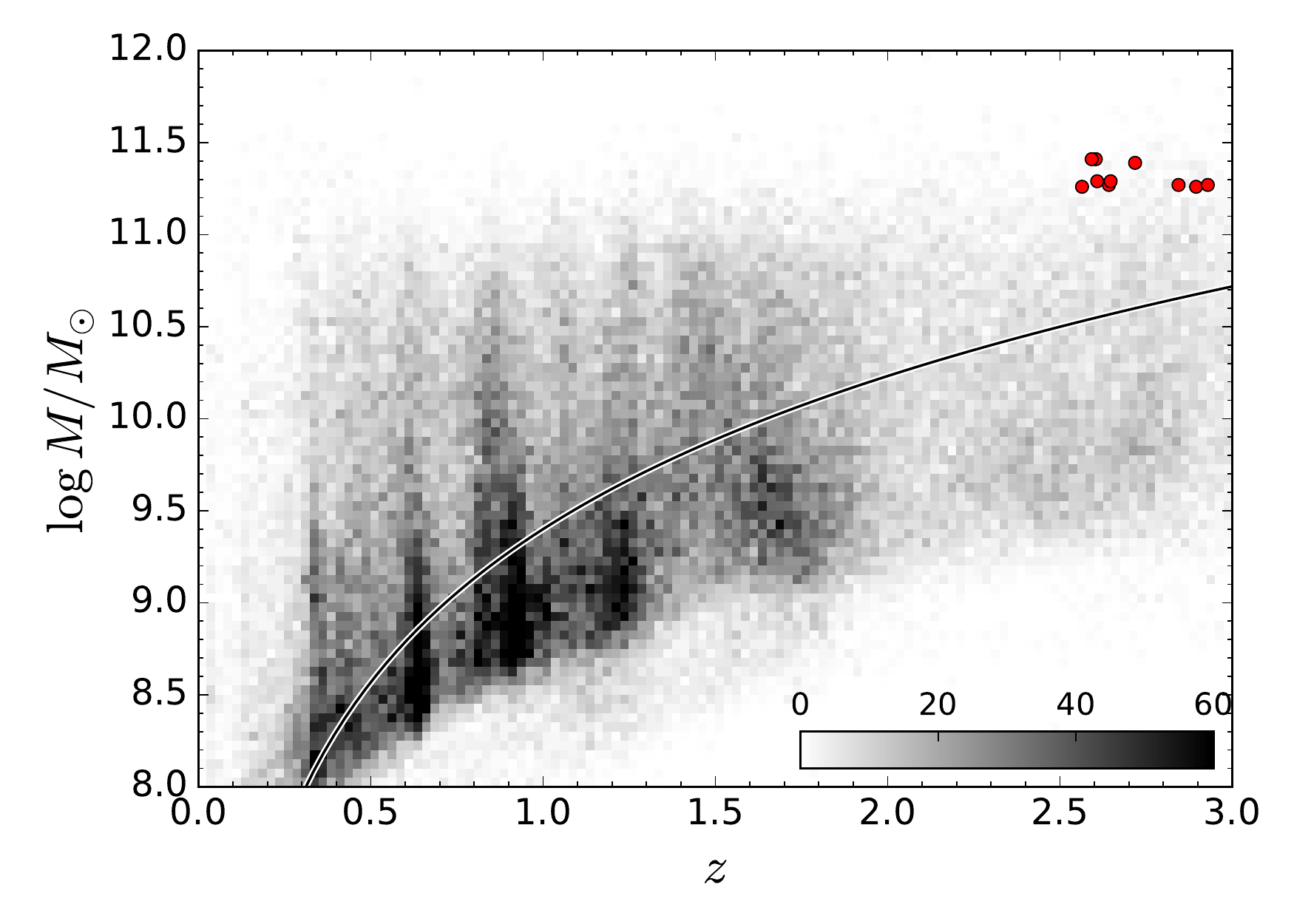}
\caption{Stellar mass vs. redshift for galaxies in the UDS with $K<24$.  The intensity of the grayscale indicates the number of galaxies in each bin.  The stellar mass completeness limit is shown by the black curve \citep{quadri2012}.  The ten QGs targeted with HST/WFC3 for our program with mass \masslimit\ at \zwindow\ are indicated in red.  These galaxies represent the extreme tail of the stellar mass function at high redshift.} \label{fig_selection}
\end{figure}

Using this catalog, we targeted the most massive QGs at high redshift.  Figure~\ref{fig_selection} shows the distribution of galaxy stellar masses with redshift for galaxies brighter than the magnitude limit of the catalog, $K<24$.  The red points represent the ten QGs above \masslimit\ at \zwindow\ that were targeted for HST/WFC3 followup observations (see Section~\ref{sec_hst}).  They clearly stand out at the extreme of the stellar mass function.  The median redshift of the sample is \zmed\ and the median stellar mass \mmed.

\begin{figure}
\epsscale{1.2}
\plotone{\plotdir 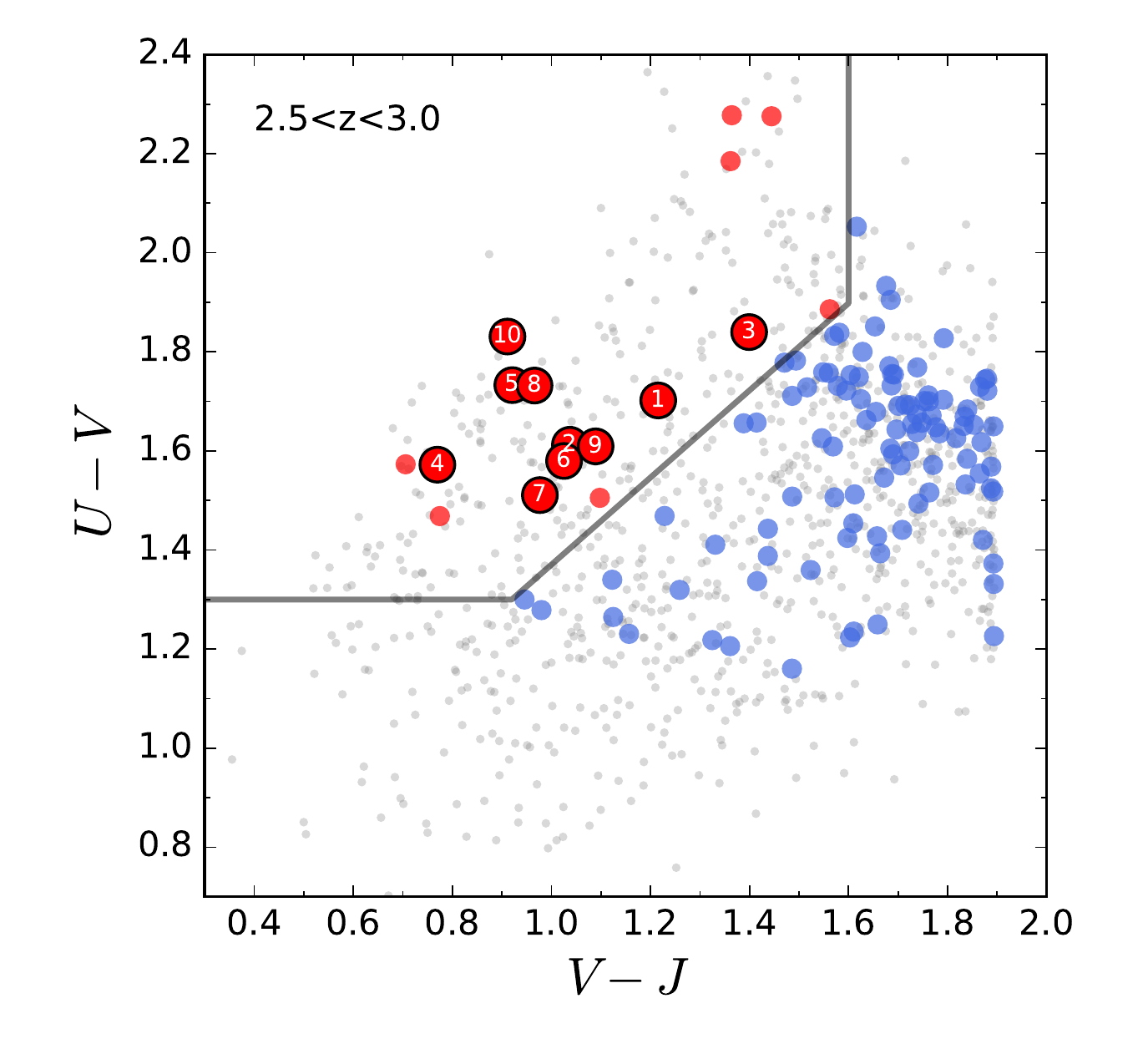}
\caption{Rest-frame $U-V$ vs. $V-J$ colors for galaxies at \zwindow\ and above $M>10^{10.7}$~\msun\ (gray).  The gray wedge represents the boundary that separates QGs and SFGs \citep{williams2009}.  Massive QGs and SFGs above \masslimit\ are indicated in red and blue, respectively, while the subset of massive QGs targeted with HST/WFC3 are shown numbered.  Massive QGs are uncommon at these redshifts, comprising only $15\%$ of the population.} \label{fig_uvj}
\end{figure}

The high redshift QGs were classified as quiescent based on their rest-frame $U-V$ vs. $V-J$ colors, as seen in Figure~\ref{fig_uvj}.  This color-color diagram is commonly used to distinguish QGs (top left) from SFGs (bottom right) \citep[][and others]{williams2009,patel2012}.  QGs above \masslimit\ are uncommon at \zwindow, making up only $\sim 15\%$ of the population while SFGs make up the rest.  The QGs targeted with HST are shown in red and numbered so that the reader can compare galaxy properties across different figures.  All of them had confident detections in $JHK$ and were not obvious blends or located near bright stars or bright foreground galaxies.

\begin{figure*}
\epsscale{1.1}
\plotone{\plotdir 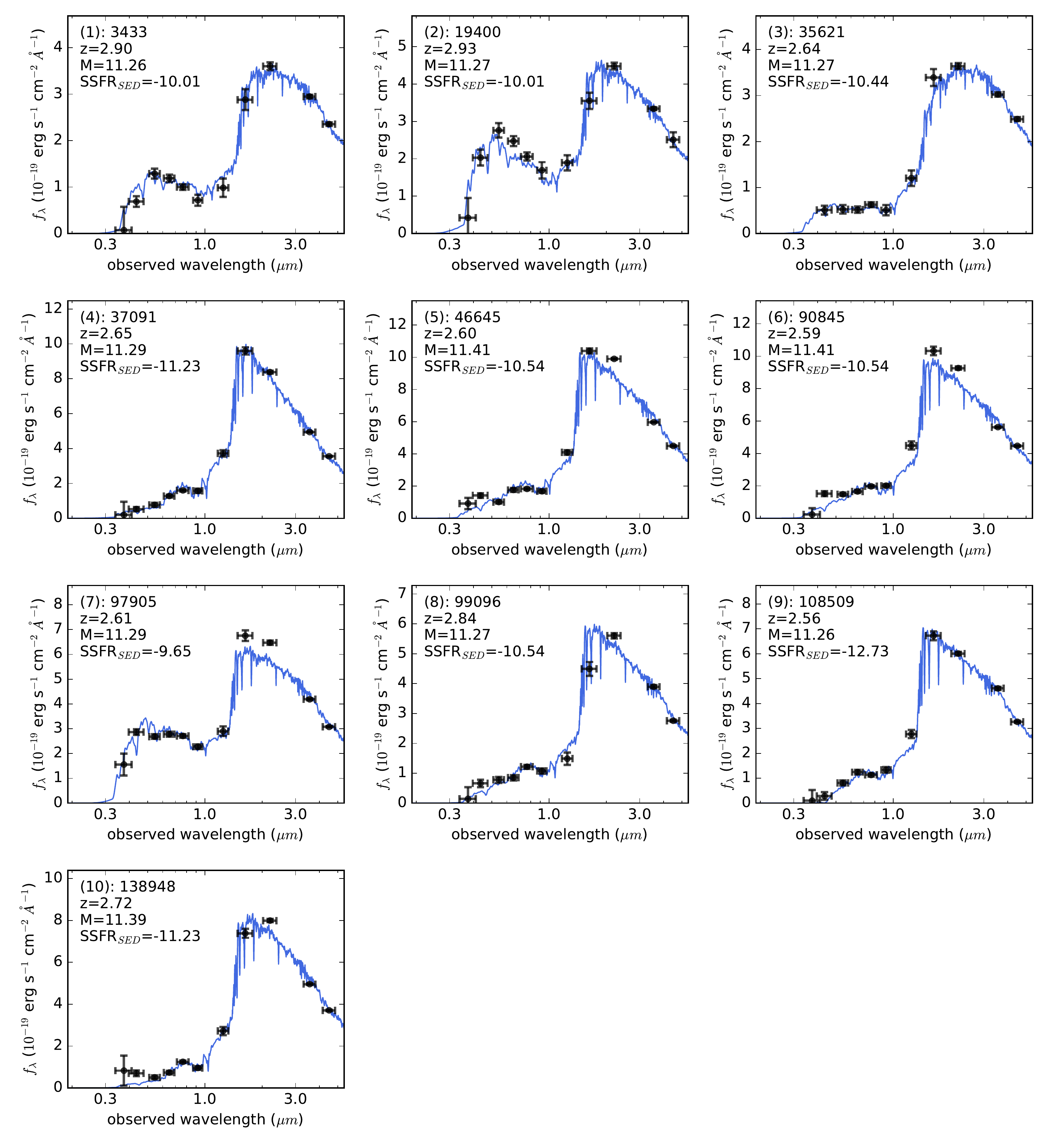}
\caption{Spectral energy distributions (SEDs) of the 10 massive QGs targeted with HST/WFC3 in the UDS.  Photometry is shown in black and the best-fitting BC03 $\tau$-model in blue (smoothed by 100~\AA).  The models fit the data well at the given photometric redshift.  These galaxies exhibit strong Balmer/4000~\AA\ breaks, indicative of intermediate age to older stellar populations and lending support to their quiescent nature.} \label{fig_sed}
\end{figure*}

Owing to their high intrinsic brightness, our HST sample of high redshift massive QGs have well constrained properties, such as their photometric redshifts and stellar masses.  Figure~\ref{fig_sed} shows spectral energy distributions (SEDs) for the targeted QGs.  The best-fitting \citet[hereafter, BC03]{bc03} $\tau$-model from FAST is plotted over the photometry, which is of high enough quality to robustly constrain both the redshift and stellar mass for each target.  All of the sources appear to have strong Balmer/4000~\AA\ breaks, indicative of intermediate age to older stellar populations.  In addition, all but one of these targets have an SED based SSFR less than $1/(3 \times t_H)$, further highlighting the degree of quiescence.

\subsection{Targeted HST WFC3 $H_{160}$ imaging} \label{sec_hst}

\begin{figure}
\epsscale{1.3}
\plotone{\plotdir 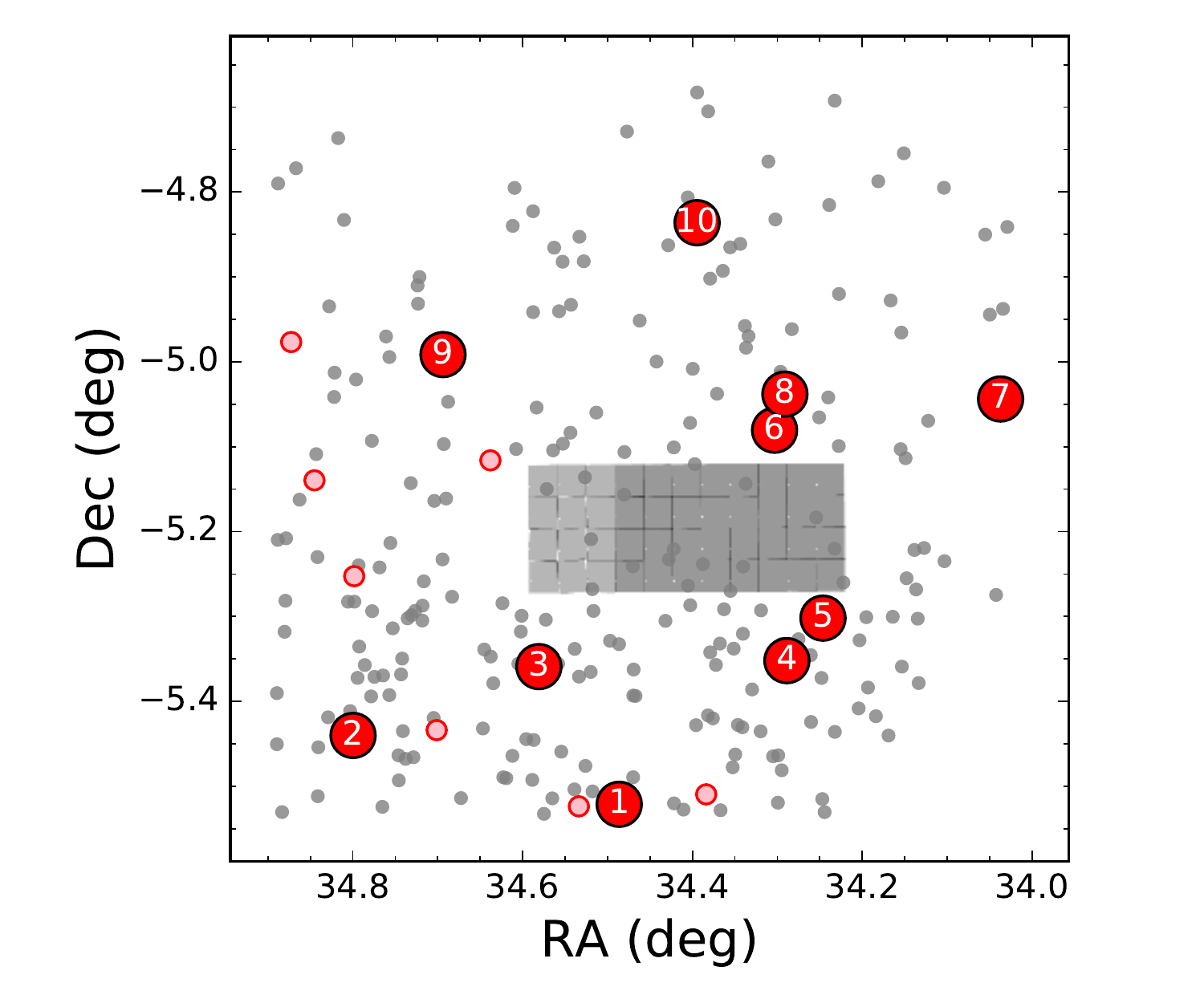}
\caption{Spatial distribution of QGs at \zwindow\ with masses above $M>10^{10.7}$~\msun\ in the UKIDSS UDS field.  Many of these galaxies lie within the CANDELS WFC3 footprint (shaded region).  However, the most massive QGs (\masslimit, colored points) lie predominantly outside of the CANDELS footprint.  The subset of these QGs that were targeted with HST WFC3 $H_{160}$ imaging are numbered.} \label{fig_map}
\end{figure}

\begin{figure*}
\epsscale{1}
\plotone{\plotdir 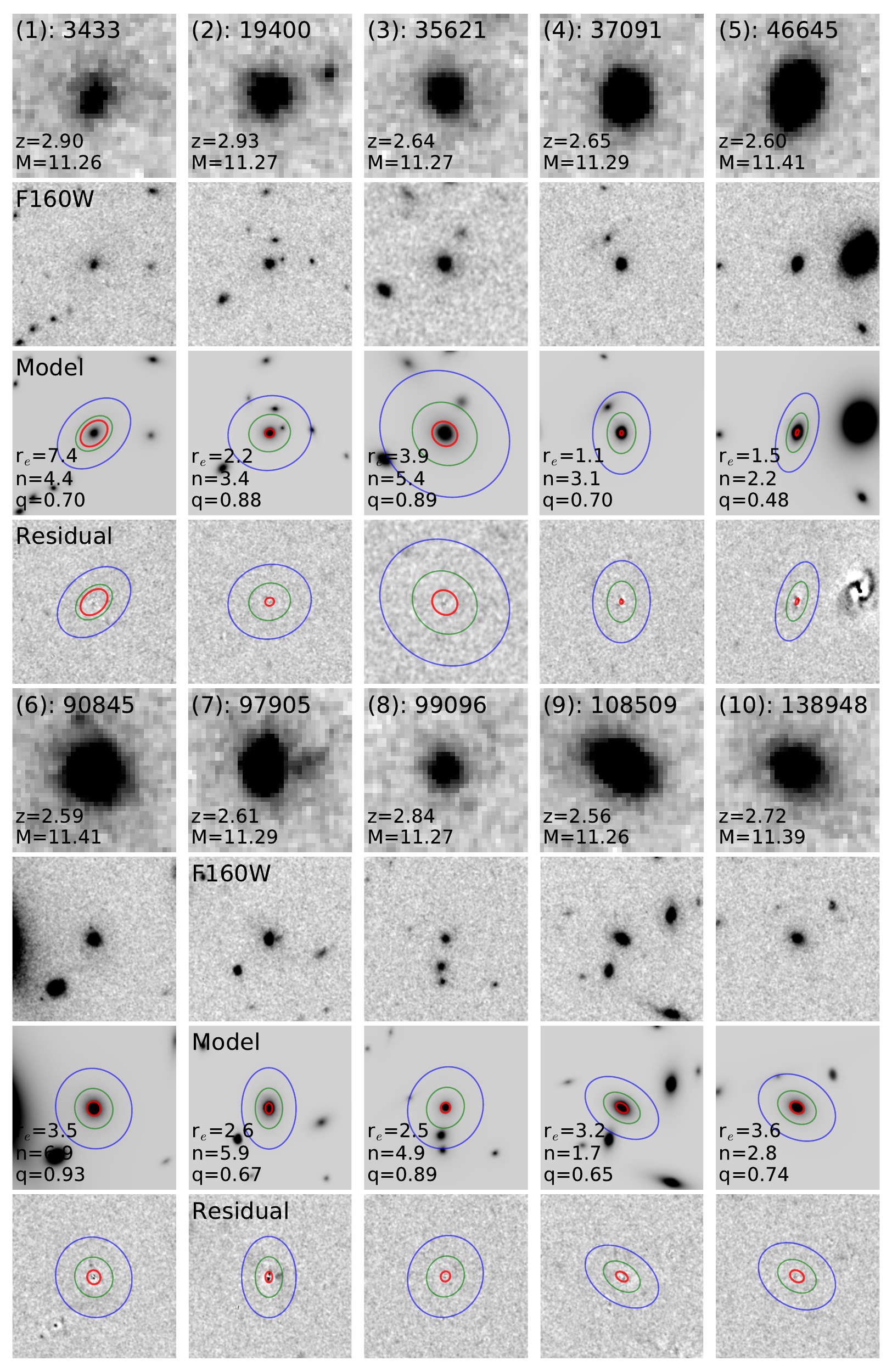}
\caption{HST WFC3 $H_{160}$ cutouts of the sample of ten massive QGs at \zwindow.  Top panels show zoomed in views of the targets ($\sim 2$\arcsec\ on a side).  The second row shows the full image that was used as input to GALFIT ($10$\arcsec\ on a side, except \#3, which was $6$\arcsec).  The third row shows the best-fit GALFIT model for one of the PSF stars along with the best-fit values for $R_e$ in kpc, Sersic $n$, and axis ratio $q$.  Note that neighbors are also modeled as part of the fit.  The half-light radii are indicated by the red ellipses while radii of $10$ and $20$~kpc are indicated in green and blue, respectively.  The bottom row shows the residual image.} \label{fig_ps}
\end{figure*}

\begin{deluxetable*}{llccccccccccc}
\tablecolumns{13}
\tablecaption{Properties of quiescent galaxies at \zwindow}
\tablehead{
  \colhead{ID} &  \colhead{UDS ID} & \colhead{RA} & \colhead{Dec} & \colhead{$z_{\rm phot}$} & \colhead{$\log M/M_{\odot}$} & \colhead{$H_{160}$}  & \colhead{$R_e$} &  \colhead{$\sigma_{R_e}/R_e$} & \colhead{$b/a$} &  \colhead{$\sigma_{(b/a)}/(b/a)$} & \colhead{Sersic} &  \colhead{$\sigma_n/n$} \\ 
  \colhead{} &  \colhead{} &  \colhead{(deg)} &  \colhead{(deg)} &  \colhead{} & \colhead{(dex)} & \colhead{(mag)} & \colhead{(kpc)} &  \colhead{} & \colhead{} &  \colhead{} & \colhead{$n$} &  \colhead{} \label{table_properties}
} 
\startdata
1 & 3433 & 34.4862 & -5.5213 & 2.90 & 11.26 & 22.93 & 7.1 & 0.30 & 0.70 & 0.10 & 4.2 & 0.17 \\
2 & 19400 & 34.7997 & -5.4402 & 2.93 & 11.27 & 22.99 & 2.0 & 0.04 & 0.90 & 0.04 & 3.4 & 0.05 \\
3 & 35621 & 34.5808 & -5.3591 & 2.64 & 11.27 & 23.00 & 3.9 & 0.19 & 0.86 & 0.07 & 5.6 & 0.13 \\
4 & 37091 & 34.2887 & -5.3521 & 2.65 & 11.29 & 21.98 & 1.1 & 0.05 & 0.69 & 0.06 & 3.3 & 0.14 \\
5 & 46645 & 34.2462 & -5.3020 & 2.60 & 11.41 & 21.73 & 1.5 & 0.03 & 0.48 & 0.06 & 2.3 & 0.09 \\
6 & 90845 & 34.3033 & -5.0805 & 2.59 & 11.41 & 21.88 & 3.4 & 0.32 & 0.90 & 0.04 & 6.8 & 0.20 \\
7 & 97905 & 34.0371 & -5.0438 & 2.61 & 11.29 & 22.08 & 2.7 & 0.16 & 0.64 & 0.03 & 6.4 & 0.19 \\
8 & 99096 & 34.2911 & -5.0381 & 2.84 & 11.27 & 23.07 & 2.4 & 0.11 & 0.91 & 0.11 & 4.9 & 0.13 \\
9 & 108509 & 34.6938 & -4.9915 & 2.56 & 11.26 & 22.16 & 3.2 & 0.02 & 0.64 & 0.02 & 1.7 & 0.05 \\
10 & 138948 & 34.3945 & -4.8361 & 2.72 & 11.39 & 22.19 & 3.7 & 0.05 & 0.73 & 0.03 & 2.9 & 0.08
\enddata
\tablecomments{The UDS ID indicates the ID in the \citet{williams2009} catalog.  The $H_{160}$ magnitude is the SExtractor MAG\_AUTO magnitude.  The half-light radius, $R_e$, reported here represents the median semi-major axis from using different PSF stars with GALFIT.  The fractional uncertainties reported for $R_e$, $b/a$, and $n$ include contributions from the sky and PSF stars.}
\end{deluxetable*}

\begin{figure}
\epsscale{1.2}
\plotone{\plotdir 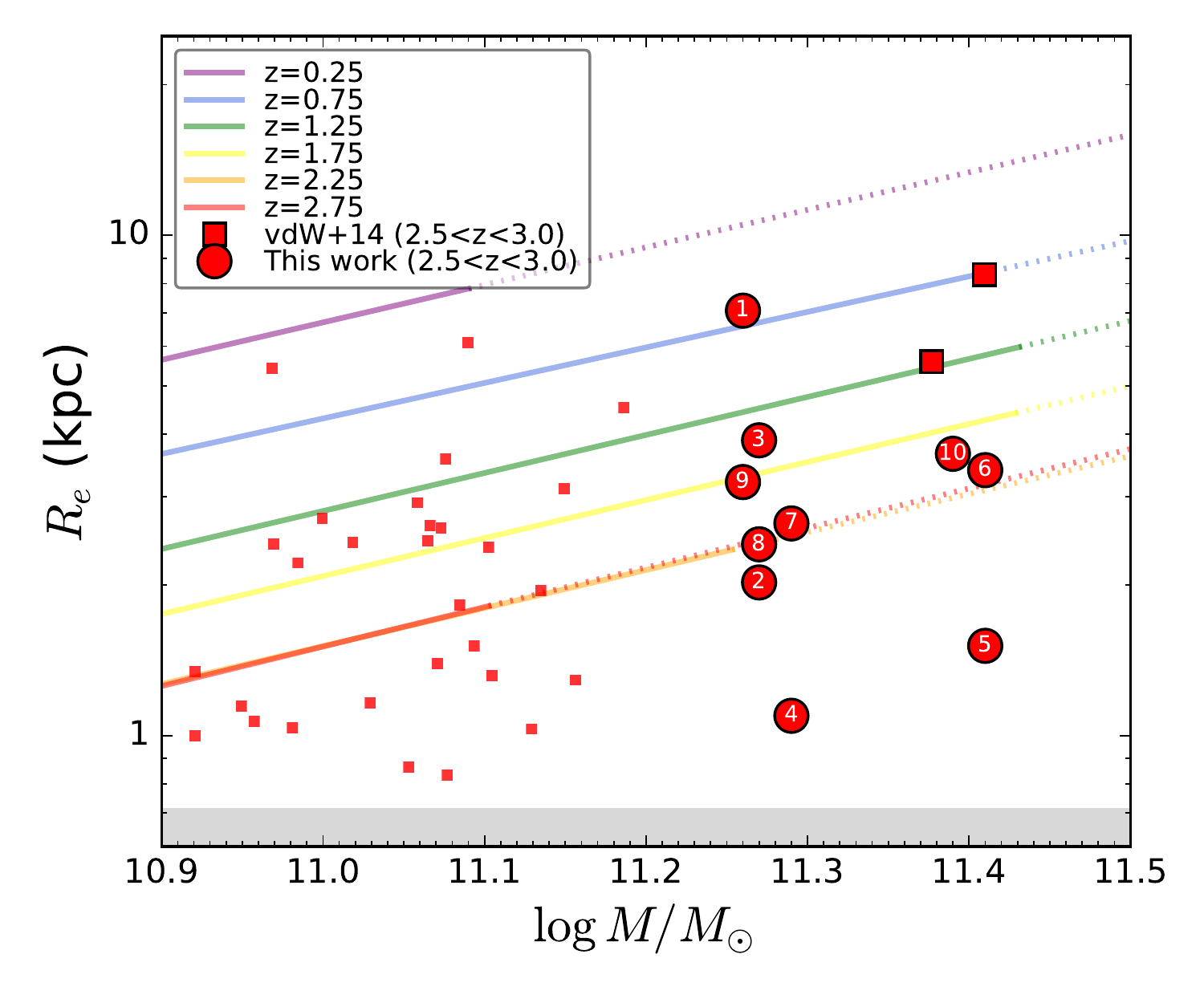}
\caption{Half-light radius vs. stellar mass.  Our UDS sample of QGs at \zwindow\ with \masslimit\ are shown as red circles and numbered.  All of the size measurements lie above the shaded gray region, which represents radii below the WFC3 $H_{160}$ resolution limit.  The sample of QGs at $2.5<z<3$ from the combined CANDELS/3D-HST fields from \vdw\ are shown as red squares, with those above our selection mass having a larger symbol size.  The colored lines represent the size-mass relations for QGs from \vdw\ (note the two highest redshift relations are similar and therefore fall on top of each other).  Extrapolations of these relations (dotted lines) begin above a stellar mass where 10 massive galaxies remain (i.e., similar to our UDS sample size).  At the highest redshifts, $z>2$, the sizes of the most massive galaxies are poorly constrained due to limited samples.} \label{fig_sizemass}
\end{figure}

\begin{figure}
\epsscale{1.2}
\plotone{\plotdir 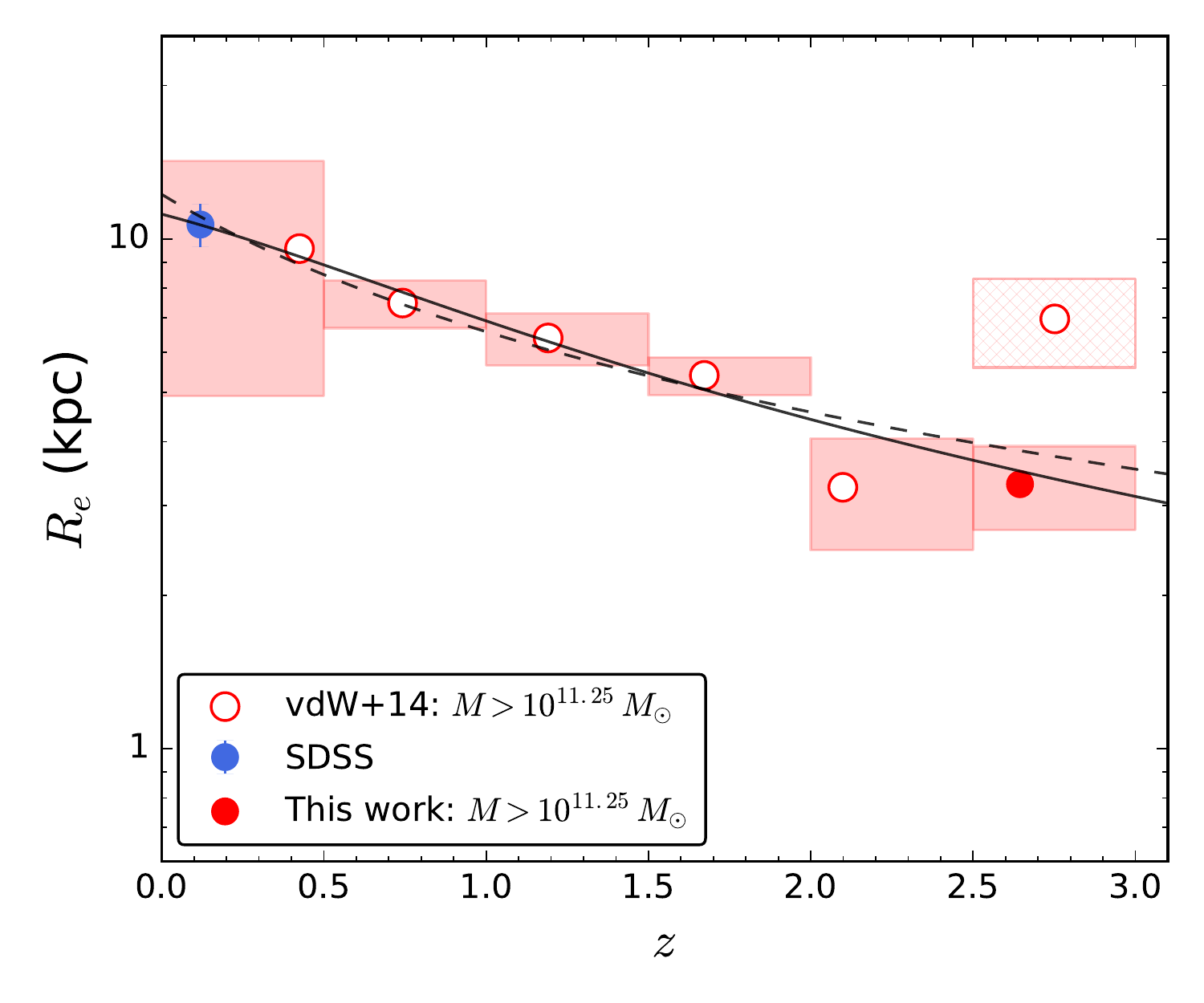}
\caption{Median half-light radius vs. redshift for QGs above \masslimit.  Measurements based on data from \vdw\ alone are shown as open red circles, while the combined sample at \zwindow\ of ten UDS QGs presented in this work and the two QGs from \vdw\ (hatched box) is represented by the solid red circle.  Also shown is a $z=0.12$ datapoint from the SDSS sample used in this work (blue).  The dashed curve represents a fit to median sizes of the form $R_e=A_z(1+z)^{\beta_z}$ while the solid curve is a fit of the form $R_e=A_H(H(z)/H_0)^{\beta_H}$.  Best fit parameters are given in Table~\ref{table_fits}.} \label{fig_sizez}
\end{figure}

In order to compare structural properties of $z \sim 3$ galaxies to those in the nearby universe at a common rest-frame optical wavelength, IR imaging is required.  The seeing of the ground-based UDS near-IR imaging (FWHM of $\sim 0\farcs8$) is insufficient for reliably measuring sizes of the smallest galaxies at $z \sim 3$.  We therefore utilize HST imaging in the reddest WFC3 band, $H_{160}$, enabling size measurements at these redshifts down to at least $R \sim 0.7$~kpc.

Despite the multitude of WFC3 imaging that covers the five CANDELS fields \citep{grogin2011,koekemoer2011} that are commonly used for high redshift studies, the space density and clustering properties of massive galaxies severely limits the number of such objects that fall within these fields.  QGs are even less common, as seen in Figure~\ref{fig_uvj}.  Figure~\ref{fig_map} shows the spatial distribution of QGs in the UKIDSS UDS field.  The lack of massive QGs in the CANDELS UDS footprint motivated our program to target a sample over a larger area.  The ten QGs that were targeted with HST/WFC3 in our Cycle 20 program (13002) and presented in this work are shown in red and numbered.  Another two QGs at $3<z<3.5$ were also targeted, however, low S/N at those redshifts led us to restrict the sample to below $z=3$.

The $H_{160}$ data were reduced with AstroDrizzle in a similar manner as CANDELS imaging \citep{koekemoer2011}.  The exposures from the 4-point dither pattern were combined to a final pixel scale of $0\farcs06$.  The total exposure time for each galaxy was $\sim 2400$~s, and double that for the two galaxies above $z>3$.  We omitted one of the dithers for QG \#1 (see Table~\ref{table_properties}) due to a nearby satellite trail, shortening its total exposure time to $\sim 1800$~s.

\subsection{SDSS: low redshift comparison sample} \label{sec_data_sdss}

We selected QGs in SDSS to serve as a low redshift comparison sample.  We used the \citet{brinchmann2004} MPA-JHU DR7 catalogs to select galaxies at $z=0.12$ ($0.118<z<0.122$) with stellar masses $10^{11.25}<M/M_{\odot}<10^{11.3}$ (median of $M=10^{11.27}$~\msun).  Their SSFRs were selected to lie below $<10^{-11}$~yr$^{-1}$.  We note that at lower redshifts than these, the outskirts of massive galaxies are heavily contaminated by interlopers posing a challenge for Sersic profile fitting.  Finally, we selected galaxies in Stripe 82, where the deepest SDSS imaging lies.  The data therefore reach low surface brightness limits and enable high fidelity measurements for structural properties.  These selection criteria led to a sample of $14$ QGs.  We used the sky-rectified $g$-band reductions of \citet{fliri2016}, which have been optimized to reach lower surface brightness limits compared to other available reductions \citep[e.g.,][]{annis2014,jiang2014}.  The $H_{160}$ light at $z \sim 3$ samples rest-frame $g$-band, which we employ for the SDSS sample.  The typical seeing in the Stripe 82 images was $\sim 1\farcs2$, which corresponds to a HWHM of $R \sim 1.3$~kpc at $z=0.12$.  All of the QGs in the SDSS sample have half-light radii well above this limit.

\subsection{Single component 2D Sersic profile fits} \label{sec_galfit}

We used GALFIT \citep{peng2002} to measure half-light radii and other structural properties based on single component Sersic fits to the 2D light distributions of our galaxies.  Initial guesses for the position, magnitude, size, axis ratio, and position angle for each galaxy were derived from SExtractor \citep{bertin1996}.  The Sersic index, $n$, was constrained to values of $0.25<n<10$.  Neighboring galaxies were also modeled and included as part of the fit.  We estimate the sky background level using the mode of sky pixel values and find this method to be more reliable than leaving the parameter free in GALFIT (see Appendix).  The uncertainty map used as input to GALFIT was constructed by adding Poisson noise to that from the sky background, as measured in an annulus around each galaxy.  The Sersic model is convolved with a PSF prior to fitting the data.  We used stars in the HST imaging, selected from color-color cuts, to serve as PSF models and ran GALFIT on each galaxy with each of the $\nstar$ PSF stars.  This provided a measure of the uncertainty in the Sersic parameters due to the PSF model.  The structural parameters reported here for a given galaxy represent the median values of the results from those PSF stars that gave valid GALFIT output.  In some cases, particular galaxy and PSF star combinations did not yield valid results, owing less to the star itself and more to the delicate nature of fitting multiple neighboring objects.  As a result, not all of our $z \sim 3$ QGs have measurements from all $\nstar$ PSF stars.  Noise from the sky background is factored into the Sersic parameter uncertainties as follows.  We take the model fit from one of the PSF stars and add to it a portion of blank sky and run GALFIT on this mock image.  This procedure is repeated with several different patches of blank sky.  The scatter in the Sersic parameters from the different mock images is then combined in quadrature with that from the PSF.

Figure~\ref{fig_ps} shows zoomed-in $H_{160}$ postage stamps for each of the high redshift massive QGs (row 1), along with the input GALFIT image (row 2), the best-fit Sersic model (row 3), and residual (row 4).  The stretch reflects $-5\sigma$ to $+20\sigma$, where $\sigma$ represents the noise in the sky background.  This figure shows the results from employing only one of the PSF stars.  In the case of \#3, a smaller cutout was used since a diffraction spike from a nearby bright star was impacting the fit.

While many studies report sizes using circularized radii, $R_e=\sqrt{ab}$, we use the semi-major axis of the half-light ellipse, $R_e=a$.  This choice allows us to compare our results directly with \vdw.  We also use their Equation~2 to correct for color gradients and report sizes that are standardized to $\lambda_{\rm rest}=5000$~\AA\ (typically a $\sim 4\%$ reduction in sizes).  We report results from GALFIT in Table~\ref{table_properties} along with other galaxy properties for our high redshift sample.

Sersic profile fitting with GALFIT was carried out in a similar manner for our SDSS sample.  For the PSF, we used the PSF stars provided by \citet{fliri2016} for each of their sky-rectified images.

\section{Results} \label{sec_results}

\subsection{The size-mass plane} \label{sec_sizemass}

Galaxy size is an important property as it reveals to first order the distribution of mass that is locked in stars.  Figure~\ref{fig_sizemass} shows the half-light radius vs. stellar mass for our massive QGs at \zwindow\ (red circles).  The median size is $R_e \sim 2.9$~kpc.  All the measurements lie well above the gray shaded region, which represents the parameter space of galaxies that would be unresolved with WFC3.  Also shown are size-mass relations for QGs from \vdw.  Their $z=2.75$ relation (shown in red) is directly comparable to our sample as the median redshifts are similar.  We note that their QGs were also $UVJ$-selected, as in our analysis, and that their stellar masses were computed  in a similar manner as in this work (see Appendix).  Two of the three massive QGs at $2.5<z<3$ from the CANDELS/3D-HST sample of \vdw\ are also shown (large red squares) and generally fall to higher $R_e$ compared to our sample and also compared to the \vdw\ $z=2.75$ size-mass relation.  As seen in Figure~\ref{fig_sizemass}, the two \vdw\ QGs lie at much higher masses compared to the rest of their QG sample (small red squares), possibly explaining their relatively larger sizes.  We have independently verified the sizes of these two QGs and note the lack of any substantial systematic differences in sizes between our work and \vdw\ (see Appendix).  A third QG from the \vdw\ sample was omitted from the analysis due to a combination of low S/N and source confusion.

The \vdw\ size-mass relations are largely extrapolated at high stellar masses: the dotted-line portion indicates where \vdw\ have samples of ten or fewer QGs (i.e., similar in number to our own sample).  For their $z=2.75$ bin ($\Delta z=0.5$), one must go to lower masses than $M=10^{11.25}$~\msun\ (i.e., our selection limit) to reach a sample of ten galaxies.  We note that their size-mass relations were weighted so that the fits would not be dominated by the more abundant low-mass galaxies.  Despite this, our massive QG sample (as well as that of \vdw) mostly lies above the extrapolated relation (i.e., dotted portion of the red line).  The median half-light radius of the combined sample of massive QGs ($R_e \sim 3.3$~kpc) lies $\sim 29\%$ above the extrapolated $z=2.75$ size-mass relation. Although this result is only significant at $1.2\sigma$ due to the small sample size, we will show in the next section that these elevated half-light radii cannot be ruled out at these high redshifts.  Interestingly, QGs at these high masses begin to deviate from the size-mass relation at $z \sim 0$ as well \citep{bernardi2011b}.

Finally, we measure an observed scatter in $\log R_e$ of $\sigma=0.25 \pm 0.04$~dex for the combined sample, which is consistent with that of lower mass QGs studied in \vdw.

\subsection{Slower size evolution for massive QGs} \label{sec_sizez}

\begin{deluxetable}{lcc}
\tablewidth{250pt}
\tablecolumns{3}
\tablecaption{Best fit parameters for \masslimit\ QG size evolution}
\tablehead{
  \colhead{$X$} &  \colhead{$\beta_{X}$} & \colhead{$\log A_{X}$} \label{table_fits}
} 
\startdata
$z$ & $-0.90 \pm 0.12$ & $1.09 \pm 0.04$ \\
$H$ & $-0.85 \pm 0.12$ & $1.05 \pm 0.04$
\enddata
\tablecomments{For $X=z$, the fitting form is $R_e=A_z(1+z)^{\beta_z}$, while for $X=H$, the fitting form is $R_e=A_H(H(z)/H_0)^{\beta_H}$.  $A_X$ is in units of kpc.}
\end{deluxetable}

\begin{figure}
\epsscale{1.2}
\plotone{\plotdir 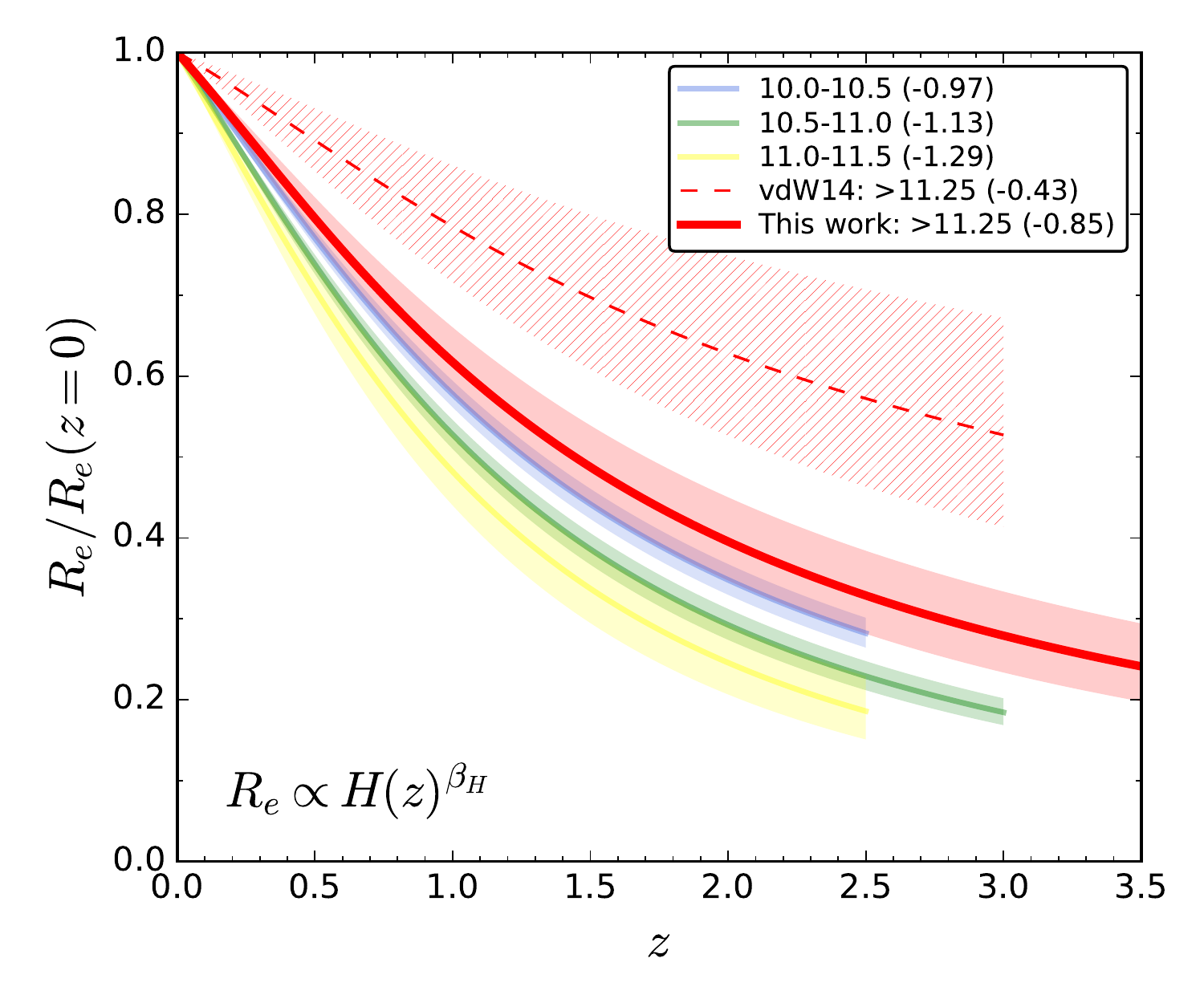}
\caption{A comparison of the best fit size evolution with $H(z)$ for QGs in different mass bins.  These are normalized by the size at $z=0$.  The shaded region reflects the $1\sigma$ uncertainty on $\beta_H$ (noted in the legend).  Our measurement for the highest mass QGs is indicated by the red curve while those at lower masses are from \vdw.  The dashed red curve and hatched region shows the result from utilizing the \masslimit\ sample in \vdw\ alone.  Our larger sample at high redshift has led to more stringent constraints for high mass QGs.  At the highest masses, size evolution is more gradual than at intermediate masses.} \label{fig_betaH}
\end{figure}

While the previous section explored our high redshift UDS QG sample in the context of the evolution of the overall size-mass relation, here we compare to galaxy samples specifically selected above our mass limit of \masslimit.  This allows us to fit for the evolution in half-light radii for such QGs and compare to QGs at lower masses.  Figure~\ref{fig_sizez} shows the median sizes and their uncertainty for galaxies selected at \masslimit\ at different redshifts.  Three different datasets are represented in this figure: (1) SDSS (blue), (2) \vdw\ (open red circles), and (3) our UDS sample at \zwindow\ combined with the three \vdw\ QGs in this high redshift bin (filled red circle).  We reiterate that in all of the three datasets, sizes were computed in a similar manner using GALFIT.  The extent of the shaded boxes indicate the redshift bin in the horizontal direction and the bootstrapped uncertainty on the median half-light radius in the vertical direction.  Note that the error for the highest redshift \vdw\ bin (hatched region) was taken to be the range between the two points in that bin.

Prior to our work, the \vdw\ sample of two QGs at $2.5<z<3$ with \masslimit\ indicated a fairly high median size of $R_e \sim 7$~kpc, hinting at mild size evolution to $z \sim 0$ for the high mass end.  Substantially increasing this sample by combining it with our HST observations of QGs at \zwindow, reveals that the typical massive QG was more compact with $R_e \sim 3.3$~kpc. 

Also shown in Figure~\ref{fig_sizez} are two fits to the median sizes of the form $R_e=A_z(1+z)^{\beta_z}$ (dashed line) and $R_e=A_H(H(z)/H_0)^{\beta_H}$ (solid line).  Table~\ref{table_fits} shows the best fit parameters for these fits.  The uncertainties were computed by bootstrapping the sample in each redshift bin and re-computing the model fit to the new median values.  Both parameterizations indicate that massive QGs were a factor of $\sim 3$ smaller in size at $z=3$ compared to today.  The small variation in the median mass of the sample in each redshift bin does not significantly impact these findings.  As also noted in \vdw, the parameterization with $H(z)$ provides a better fit to the data compared to the more often used $(1+z)$.  This is a result of the slower evolution of $H(z)$ at late times.

We compare the size evolution for our massive QGs to those from lower mass bins using data from \vdw\ in Figure~\ref{fig_betaH}.  The best fit size evolution curves, parameterized by $H(z)$, are normalized by the $z=0$ size.  The shaded regions reflect the $1\sigma$ uncertainty on the slope, $\beta_H$.  More negative values of $\beta_H$ indicate faster size evolution.  \vdw\ measure $\beta_H= -0.97 \pm 0.05$, $-1.13 \pm 0.06$, and $-1.29 \pm 0.16$ in mass bins (in units of $\log M/M_{\odot}$) spanning $10.0$ to $10.5$, $10.5$ to $11.0$, and $11.0$ to $11.5$, respectively.  We note that their lower mass bin (10-10.5) spans a region where the size-mass relation breaks to a much shallower slope toward lower masses \citep[e.g.,][\vdw]{mosleh2013} and should therefore be compared to with caution.  At higher masses, whether a trend of faster size evolution exists is unclear given that the values of $\beta_H$ for their two higher mass bins are within $\sim 1\sigma$ of each other.  Note also that while the boundary of the highest mass bin reported by \vdw\ extends to $10^{11.5}$~\msun, the median mass in that bin is less than our mass limit of $10^{11.25}$~\msun\ owing to the steepness of the stellar mass function.  

Our measurement at \masslimit\ (red curve), with $\beta_H=\betaH$, results in more gradual size evolution compared to the intermediate mass galaxies studied by \vdw.  Re-fitting only the high mass \vdw\ data from Figure~\ref{fig_sizez} (i.e., open circles in that figure), results in an even shallower $\beta_H=-0.43 \pm 0.16$ (dashed red line).  The constant or perhaps even declining value of $\beta$ with mass therefore does not persist to the highest masses.  The larger sample of massive QGs at \zwindow\ studied here has led to more stringent constraints on $\beta_H$ and we find that their sizes do not evolve as strongly as QGs at lower masses.  Employing the size evolution parameterized by $(1+z)$ does not change these conclusions.  The evolution of the size-mass relation shown in Figure~\ref{fig_sizemass} would not reveal this differential evolution with mass on its own owing to the similar slopes measured at different redshifts.  However, the median sizes of massive QGs relative to those inferred from the size-mass relations indicates an increasing ratio toward high redshift (and more so than that of lower mass QGs).  This discrepancy likely arises from the subtleties of the size-mass relation fitting carried out by \vdw.  However, just as they find differences in $\beta_H$ between their lower and higher mass bins (despite a roughly constant size-mass slope at different redshifts), we find further differences in $\beta_H$ for the highest mass QGs.

\subsection{Axis ratio distribution: the most massive QGs at $z \sim 3$ are likely spheroids} \label{sec_q}

\begin{figure}
\epsscale{1.2}
\plotone{\plotdir 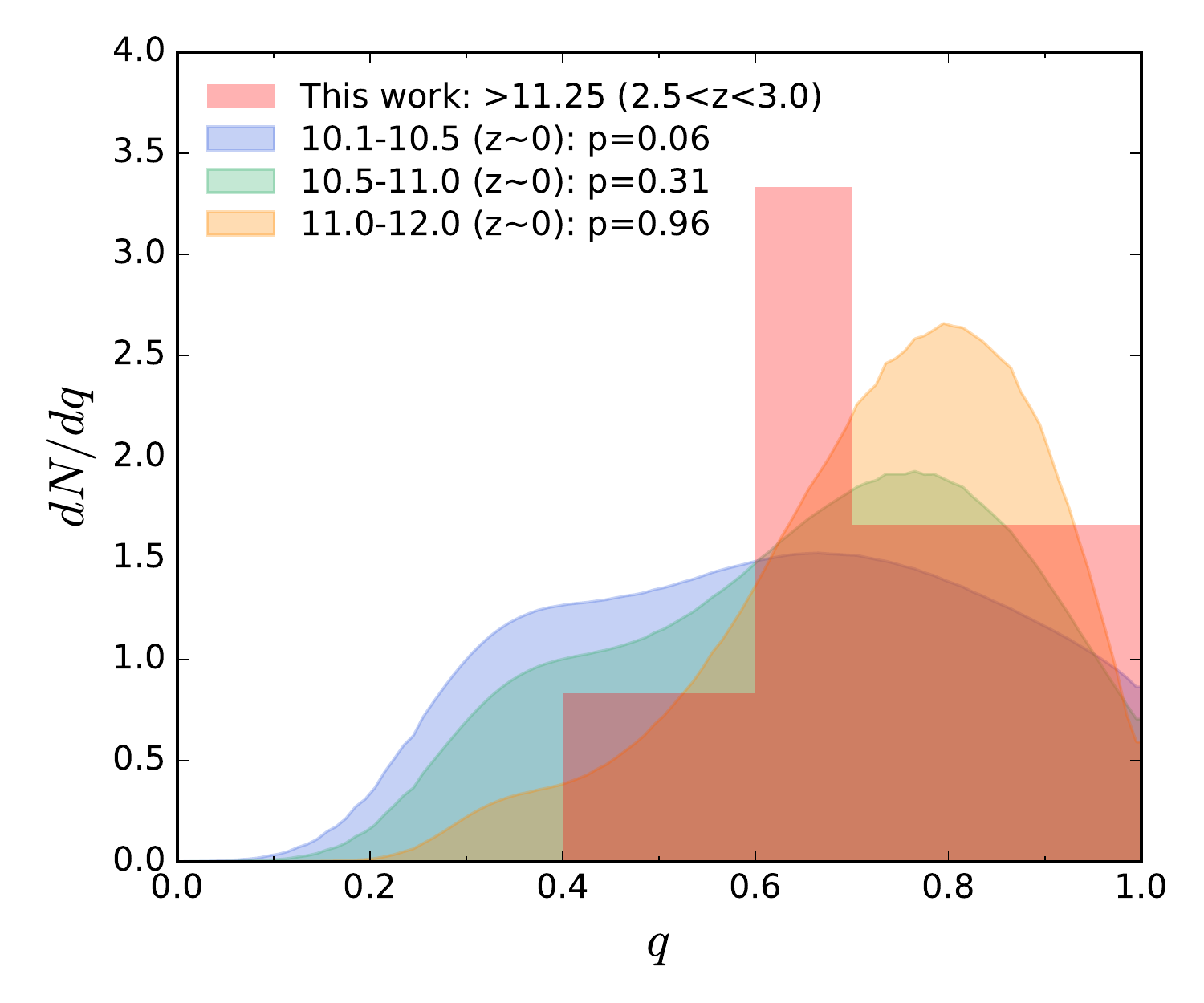}
\caption{Distribution of axis ratios ($q=b/a$) for massive QGs.  Our combined sample at \zwindow\ is shown in red in bins of $\Delta q=0.1$.  Distributions for low redshift QGs from SDSS with masses $10^{10.1}<M/M_{\odot}<10^{10.5}$ (blue), $10^{10.5}<M/M_{\odot}<10^{11}$ (green), and $10^{11}<M/M_{\odot}<10^{12}$ (orange) are also shown \citep{holden2012}.  Their analysis indicates that the most massive QGs at low redshift (orange) generally lack disks while such systems are more common at lower masses.  Two-sided KS tests indicate that our high redshift distribution is most consistent with the high mass sample at $z \sim 0$.
} \label{fig_q}
\end{figure}

\begin{figure*}
\epsscale{1.1}
\plotone{\plotdir 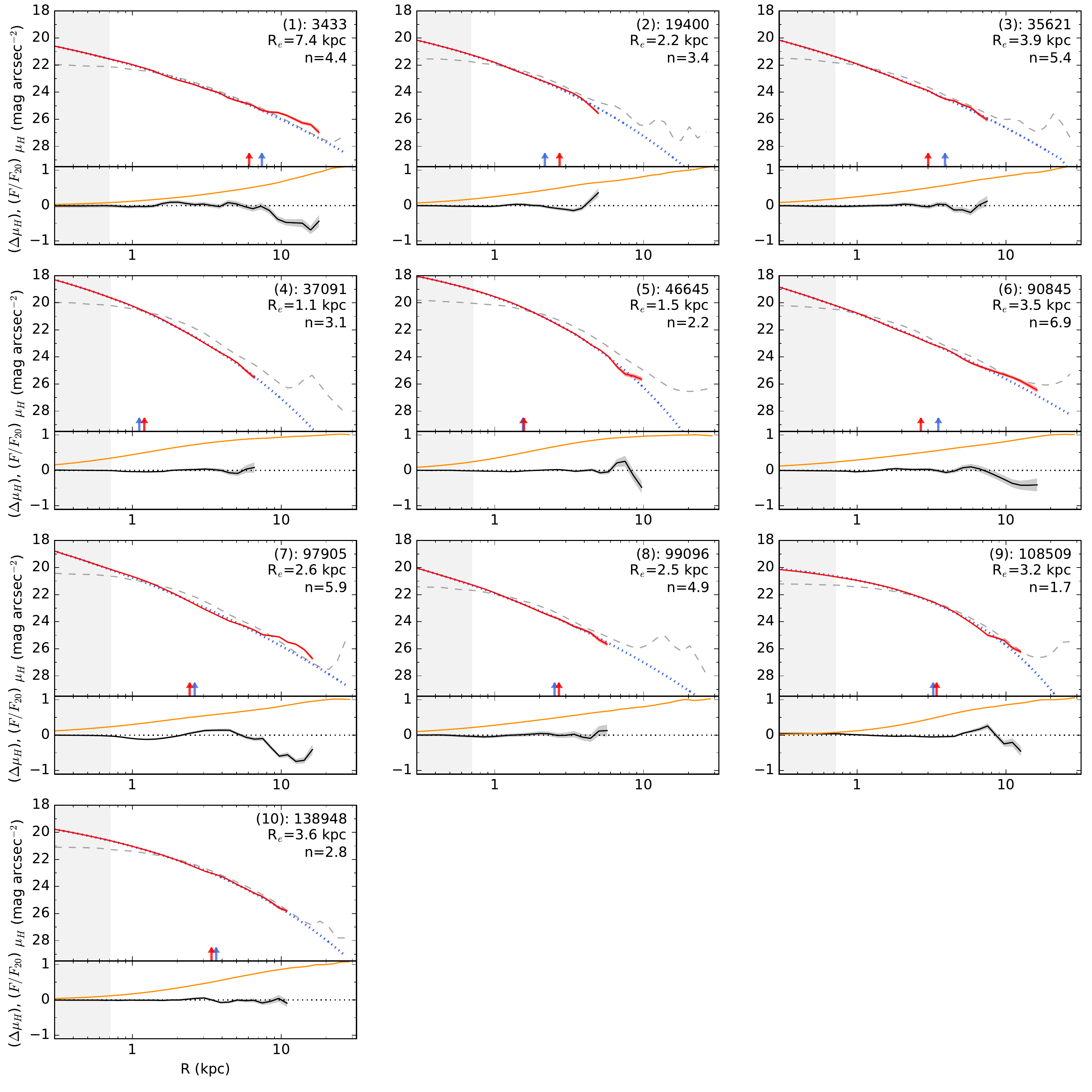}
\caption{Surface brightness profiles of $z \sim 3$ massive QGs in the observed WFC3 $H_{160}$ band.  The gray dashed curve represents elliptical aperture photometry performed on the PSF convolved best-fit GALFIT model (e.g., row 3 of Figure~\ref{fig_ps}).  Contributions from neighboring galaxies are included in this curve, hence the fluctuations at large radii in some cases.  The blue dotted curve is the deconvolved version of the best-fit Sersic profile, with neighboring galaxies subtracted out.  Finally, the red solid curve is the residual-corrected, deconvolved profile.  The bottom panels show the residuals (black curve) as well as the cumulative flux normalized by the value at $R=20$~kpc ($F_{20}$, orange).  The blue arrow indicates the half-light radius from the best-fit GALFIT model while the red arrow indicates the half-light radius from the residual-corrected profile and assuming $F_{20}$ approximates the total flux.  The gray shaded region represents the radial region encompassing the HWHM of the PSF.  The residual corrected profiles, which can reveal departures from a Sersic parameterization, in fact show that these QGs follow such profiles out to multiple effective radii.  Fluctuations from a pure Sersic profile are typically only seen at large radii where residuals from neighboring galaxies may play a role.  The profiles shown here are based on only one of the PSF stars used with GALFIT.} \label{fig_profiles}
\end{figure*}

While the sizes discussed in previous sections indicate the extent of the stellar mass distribution, the axis ratio distribution provides insight into the typical intrinsic shape of that distribution (e.g., disky vs. spheroid).  Axis ratios are also one of the more robust quantities from the Sersic profile fitting.  Figure~\ref{fig_q} shows the distribution of axis ratios, $q$, for our sample of ten high redshift QGs in the UDS and two QGs from \vdw\ (red).  The mean for our sample is $q=0.74$, a high value indicating fairly spheroidal systems.  Moreover, only one out of the twelve galaxies (\#5) has $q<0.5$, highlighting the general lack of a significant tail toward lower axis ratios and therefore suggesting a dearth of disk-dominated systems.

We compare our axis ratios to the distributions from \citet{holden2012} for QGs in SDSS at $z=0.06$.  In this way, we can determine whether our high redshift massive QGs resemble those in the nearby universe.  We use the \citet{holden2012} SDSS sample rather than the one from Stripe~82 employed earlier because the former is much larger.  However, we note that a KS test indicates that the Stripe~82 sample is consistent with being drawn from the highest mass bin from \citet{holden2012}.  The axis ratio distributions for $z=0.06$ QGs for three different stellar mass bins are shown in the figure.  Each of these represent the best fit model to the SDSS data for a given mass bin.  The model was comprised of two components: (1) a triaxial component representative of spheroids, and (2) an oblate spheroid component, which is meant to mimic disks with bulges.  \citet{holden2012} find that QGs in their highest mass bin ($10^{11}<M/M_{\odot}<10^{12}$) are overwhelmingly dominated by the triaxial component, in contrast to lower mass bins where an emerging tail to low axis ratios demands a larger disky population.  A two-sided KS test with our high redshift QGs and those in the most massive SDSS bin (orange) gives a $p$-value of $0.96$ indicating that we cannot reject the null hypothesis that the two samples are drawn from the same distribution.  Meanwhile, the $p$-values are much lower for the lower mass bins as indicated in the figure.  Drawing random samples of twelve galaxies at a time from these two lower mass distributions, we find that the mean value of $q$ exceeds that of our sample in $3\%$ and $\sim 9\%$ of cases for the successively higher mass bins.  For the highest mass bin it is much higher, at $\sim 47\%$.   The small number statistics make it difficult to confidently rule out agreement with the two lower mass bins from \citet{holden2012}.  However, the excellent agreement with their highest mass bin suggests that spheroids may be dominant in the high redshift massive QG population, just as they are in their counterparts in the nearby universe.

\subsection{Surface brightness profiles} \label{sec_profiles}

\subsubsection{High redshift QGs closely follow Sersic parameterization: unlikely to be missing excess light}

The Sersic representation for the light profiles of our galaxies are idealized.  There is no reason to assume a priori that high redshift galaxies are well described by such profiles.  For example, a galaxy with a diffuse extended disk within a more dominant bulge might deviate from a single Sersic profile and bias the half light radius.  Here, we measure residual-corrected light profiles following the work of \citet{szomoru2010,szomoru2011,szomoru2012,szomoru2013} in order to investigate whether our galaxies are well-modelled by single component Sersic profiles and their corresponding half-light radii.

The residual-corrected profiles are measured as follows: (1) the best-fit parameters from the GALFIT Sersic fit are used to construct a deconvolved 2D Sersic model, (2) photometry is performed in elliptical apertures on both the deconvolved Sersic model and the residual image using shape parameters from the GALFIT best-fit model, (3) the residual profile is added to the deconvolved model profile to complete the residual-corrected profile.  The intent of this procedure is similar in nature to CLEAN \citep{hogbom1974}, in which the residuals are added to a deconvolved map.  In addition to preserving flux, a first order correction is applied to the assumed intrinsic profile.

Figure~\ref{fig_profiles} shows surface brightness profiles for our high redshift sample in the observed WFC3 $H_{160}$ band.  These results are based on one of the PSF stars but are generally representative of our overall findings.  Note that the radius here indicates the semi-major axis of the elliptical aperture.  The gray dashed curves represent aperture photometry carried out on the best-fit GALFIT image (e.g., 3rd row of Figure~\ref{fig_ps}).  These are not deconvolved for the PSF, hence the flattening within the PSF HWHM (gray shaded region).  In some cases, these gray curves fluctuate at large radii, indicating the presence of un-subtracted neighboring galaxies.  The dotted blue curves show the profile of the deconvolved best-fit Sersic model of the primary galaxy alone (i.e., neighbors are not included here).  Note that these deconvolved profiles reveal the steeper cores that are smoothed out by the WFC3 PSF.  Finally, the residual-corrected profile is indicated by the solid red curve and the $1\sigma$ uncertainty by the shaded red region.  The difference between the residual-corrected profile and the deconvolved model are shown in the bottom panels.  Also shown in this panel is the cumulative flux (orange curve), as measured from the residual-corrected profile and normalized by the flux at $R=20$~kpc ($F_{20}$).

Surface brightness limits were computed by placing boxes across regions of sky and determining the scatter in total flux.  This process was repeated for boxes with varying areas allowing us to determine $3\sigma$ limits for elliptical annuli with matching areas.  At each radius, we measure the surface brightness limit based on the corresponding annular aperture area used to compute $\mu_H$.  We truncate the residual-corrected profiles at radii where the surface brightness limit exceeds the residual-corrected profile.  In general, we are able to measure profiles to depths of $\sim 26$~mag~arcsec$^{-2}$, which corresponds to roughly $\sim 10$~kpc for most of the sample.

The residual-corrected profiles reveal several key aspects about the nature of massive QGs at high redshift.  Most importantly, their surface brightness profiles follow a Sersic parameterization out to several half-light radii.  This is clearly seen in the figure by how closely the residual-corrected profiles (solid red curve) follow the deconvolved Sersic models (dotted blue curve).  Deviations from a Sersic profile at large radii typically occur in regions where residuals from neighboring galaxies impact the sky level.  These regions are easily identified by the fluctuations in the gray dashed line.  Only in one of the ten galaxies, \#1, does it definitively appear that excess light exists on top of the Sersic profile at large radii (e.g., $R \sim 10$~kpc).  The residual-corrected profiles for all of the other galaxies show that we are not missing any excess low surface brightness light out to several half-light radii.  As a consequence, the inferred half-light radii from these residual-corrected profiles (red arrow) closely match the results from the single component Sersic fit from GALFIT (blue arrow).  The scatter between the two is $\sim 0.1$~dex and the offset is negligibly small.  The sizes of these high redshift massive QGs are therefore robustly measured and truly compact, as we discuss in the next section.

\begin{figure*}
\epsscale{1.2}
\plotone{\plotdir 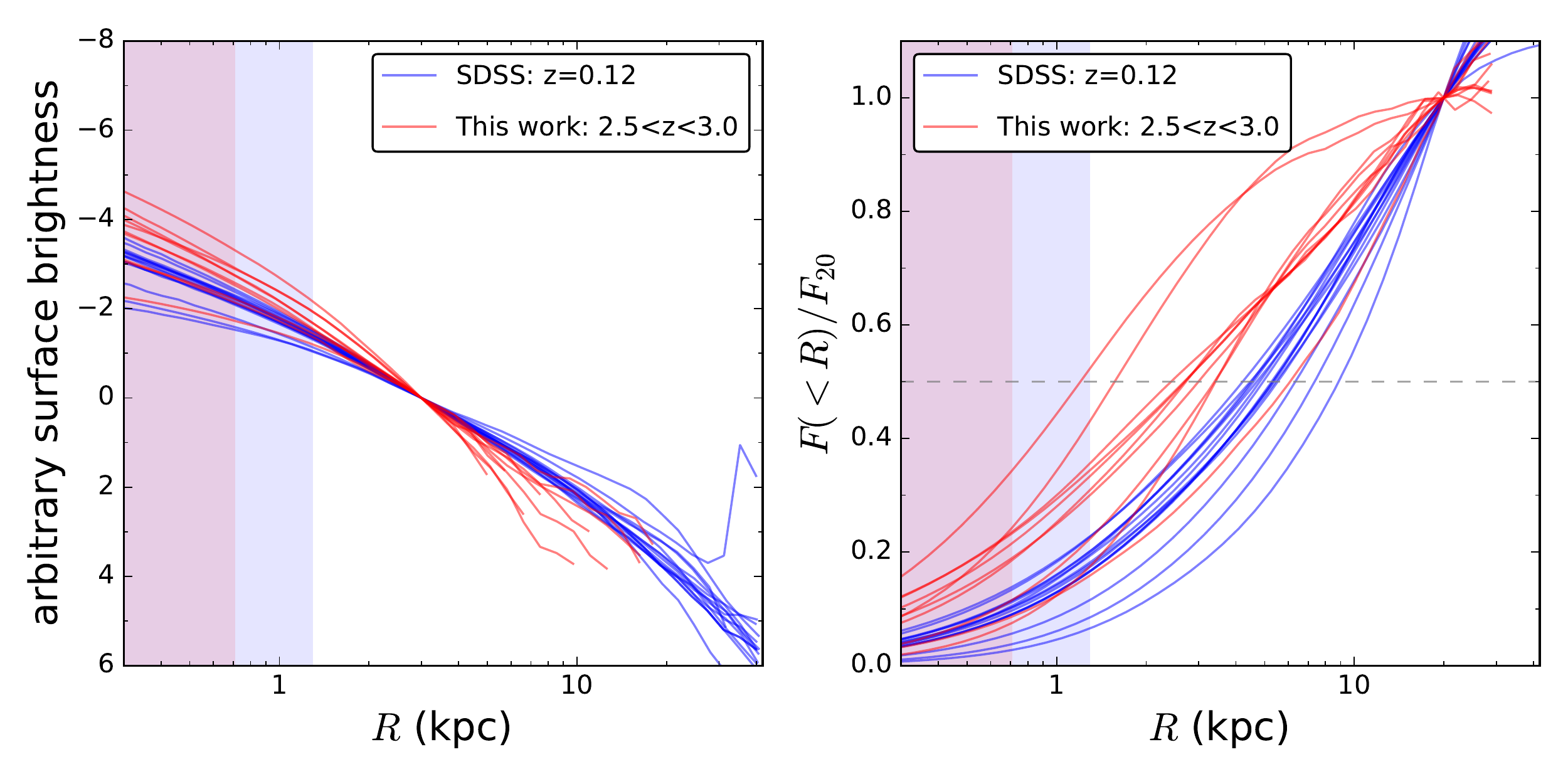}
\caption{($a$) Comparison of residual-corrected surface brightness profiles of our high redshift UDS sample of QGs (red) and low redshift SDSS Stripe 82 sample (blue).  Both samples have a median stellar mass of \mmed.  SDSS profiles are measured in the $g$-band, which is close to the rest-frame of the WFC3 $H_{160}$ imaging for our $z \sim 3$ galaxies.  The HWHM of the SDSS and WFC3 imaging are indicated by the shaded blue and red regions, respectively.  The profiles are normalized to have the same surface brightness at $r=3$~kpc.  At smaller radii the high redshift galaxies typically have steeper profiles while at larger radii their profiles fall below that of their low redshift counterparts. ($b$) Cumulative flux normalized by the flux at $R=20$~kpc.  The typical half-light radius is much smaller for the high redshift QGs compared to their low redshift counterparts.  Given the more extended nature of low redshift QGs, their true half-light radii are even larger than what is portrayed by the chosen normalization radius shown here.  Nevertheless, this figure shows that massive QGs were more compact at high redshift.} \label{fig_compare}
\end{figure*}

\subsubsection{Comparison to QGs in SDSS Stripe 82}

We compare the residual-corrected profiles of our high redshift massive QGs to those in the nearby universe from SDSS in Figure~\ref{fig_compare}.  The median stellar mass of each sample is the same.  The SDSS profiles were constructed in the same way as our high redshift profiles.   They are truncated at the limiting surface brightness depth ($\mu_g \sim 27-28$~mag~arcsec$^{-2}$), which often leads to profile measurements beyond $R \sim 40$~kpc.  As noted in Section~\ref{sec_data_sdss}, the $g$-band light from the SDSS imaging traces similar rest-frame wavelengths as the $H_{160}$ imaging for our $z \sim 3$ galaxies. 

The left panel shows that the profiles of high redshift QGs (red) are more centrally concentrated compared to low redshift QGs (blue).  Owing to their larger half-light radii and higher Sersic indices (median $n \sim 5.2$), the $z=0.12$ QGs have more extended profiles.  Further emphasizing this point, the panel on the right shows the cumulative flux, normalized by the flux within $20$~kpc ($F_{20}$).  Note that the cumulative flux is known to higher significance than the differential flux shown in the first panel.  The radius at which half of the $F_{20}$ flux (horizontal dashed gray line) is reached is on average much smaller for the high redshift QG sample.  Given that a $R=20$~kpc aperture generally underestimates the total flux for the SDSS galaxies, the true difference in half-light radius between the two samples is even larger than what is portrayed by this figure.  Finally, we note that some overlap does exist between the two samples in that the measured profiles of some high redshift QGs are similar to that of low redshift QGs.

\section{Discussion}  \label{sec_discussion}

We have robustly measured half-light radii for our HST/WFC3 targeted sample of massive QGs at \zwindow.  The residual-corrected profiles discussed in Section~\ref{sec_profiles} confirm the compactness of these galaxies.  Furthermore, our comparison to \vdw\ is well-grounded given that we have taken care to minimize systematics between the two data sets.  We have measured properties based on the SED, such as stellar masses and rest-frame colors, with the same methods (see Appendix for a comparison of stellar masses and half-light radii).  The half-light radii have also been measured with the same techniques.

While we find the half-light radii of QGs with \masslimit\ at $z \sim 3$ to be a factor of $\sim 3$ smaller compared to similar mass QGs at $z \sim 0$ (e.g., Figure~\ref{fig_sizez}), this degree of size evolution is not as pronounced as that of QGs below these masses.  This was shown in Figure~\ref{fig_betaH}, where size evolution fits of the form $R_e \propto H(z)^{\beta_H}$ indicate stronger evolution for lower mass QGs.  Put differently, compared to the highest mass QGs at high redshift, lower mass ones are relatively more compact compared to their $z=0$ counterparts.  We reiterate that this signal is also seen in the \vdw\ data alone, although with less significance owing to their smaller sample of massive QGs at high redshift (see Section~\ref{sec_sizez}).  We also emphasize that the mass limited samples selected and studied in this work are not meant to directly probe the evolution of descendant galaxies to low redshift as galaxies are expected to gain mass over cosmic time.  Instead, we are making the point that QGs above \masslimit\ experience slower size growth compared to lower mass ones.

What does this differential size evolution with stellar mass imply about the evolution of the most massive QGs?  Dry minor mergers have been proposed to lead to an increase in QG sizes \citep[e.g.,][]{bournaud2007c,naab2009b}.  If this is indeed a dominant mechanism, one possibility is that massive QGs have undergone accelerated size growth by the epoch of observation due to such mergers, perhaps owing to an over-dense environment.  Based on virial arguments \citep[e.g.,][]{bezanson2009}, a single 1:7 merger could account for the $\sim 30\%$ larger sizes (relative to the size-mass relation) at $z \sim 3$ seen in Figure~\ref{fig_sizemass}.  Continuing to low redshift, a smaller proportion of the mass growth for massive QGs would have to be acquired through the dry minor merger channel compared to lower mass QGs.  Such a scenario is required to match the observations of more gradual size evolution for the former.

Another intriguing clue may reside in the axis ratio distribution.  We find that the most massive QGs at high redshift are possibly spheroids as opposed to disks (Section~\ref{sec_q}).  This conclusion is based on comparisons to results at $z \sim 0$, which \citet{holden2012} find also persist at $z \sim 1$.  Continuing this trend to intermediate redshifts, $z \sim 2$, \citet{chang2013} also report that the highest mass QGs ($M>10^{11.3}$~\msun) are deficient in low axis ratio systems.  These results therefore bolster our case at $z \sim 3$ for spheroidal QGs.  However, at lower masses at these high redshifts, disks are found to be more common among the QG population \citep{vanderwel2011,chang2013b}.  If dry minor mergers drive size growth and can also simultaneously destroy disks \citep[e.g.,][]{bournaud2007c}, it could reinforce the scenario above as it would explain the lack of flatter systems in our high redshift massive QG sample as well as the more gradual pace of size evolution at $z \lesssim 3$ compared to lower mass QGs.  The latter, which are found to be disky at high redshift, would need to have their sizes grow at a relatively quicker pace (i.e. more negative values of $\beta_H$ or $\beta_z$) and ultimately end up as round systems to match the observations at $z \sim 0$.  The dry minor merger scenario is complicated, however, by the fact that QGs are rare at high redshift and that dynamical friction timescales are longer for higher mass ratios.  We should therefore leave open the possibility of a qualitatively different growth mode for the highest mass QGs.  For example, since changing galaxy populations potentially contribute toward size evolution, another possible explanation for a differential size growth with mass is a differential quenching rate of SFGs.

Finally, while our HST program has greatly increased the sample size of massive QGs at high redshift, even larger samples would be useful for confirming the conclusions presented here based primarily on $\sim 12$ QGs at \zwindow.  This is especially crucial, given the difference in evolution between intermediate and high mass QGs discussed here.

\section{Summary}  \label{sec_summary}

We have presented an analysis of the structural properties of the highest mass (\masslimit) quiescent galaxies (QGs) at \zwindow.  These galaxies are rare in the distant universe leaving even the largest surveys \citep[e.g.][\vdw]{vanderwel2014} without a clear picture for their typical structural properties.  For this reason, we used HST WFC3 $H_{160}$ imaging to target ten new QGs in the UDS, and combine them with two existing massive QGs from the CANDELS fields.  The primary conclusions of our analysis are highlighted below:

\begin{enumerate}
\item Fitting the $H_{160}$ images of these $z \sim 3$ QGs with 2D Sersic profiles, we measure a median half-light radius of $R_e\sim 3$~kpc, which is a factor of $\sim 3$ smaller than similar mass galaxies in the nearby universe.
\item We measure the median sizes of QGs above \masslimit\ to evolve as $R_e \propto H(z)^{\betaH}$ (or $R_e \propto (1+z)^{\betaz}$).  This indicates more gradual evolution compared to QGs below these masses which tend to have $\beta_{H} \lesssim -1$.
\item The residual-corrected surface brightness profiles of these massive QGs indicate that the Sersic parameterization is an accurate representation of the light profile out to several effective radii and that any excess light is not unaccounted for at these distances.  The profiles therefore confirm the compact nature of high redshift massive QGs when compared to similar mass galaxies in SDSS.  
\item We compare the axis ratio distribution from our $z \sim 3$ massive QGs to that in the nearby universe and find that the high redshift QGs are most consistent with distributions that are not disk dominated.
\end{enumerate}

\acknowledgements
We thank the anonymous referee for their helpful feedback.  We are extremely grateful to Daniel Kelson for providing detailed feedback and suggestions that greatly improved the manuscript.  We also thank John Mulchaey for his feedback.  Support for program \#13002 was provided by NASA through a grant from the Space Telescope Science Institute, which is operated by the Association of Universities for Research in Astronomy, Inc., under NASA contract NAS 5-26555.  Y. X. H. acknowledges support from the Rose Hills Foundation.

\appendix

\section{Comparison with \vdw}

\begin{figure}
\epsscale{1.2}
\plotone{\plotdir 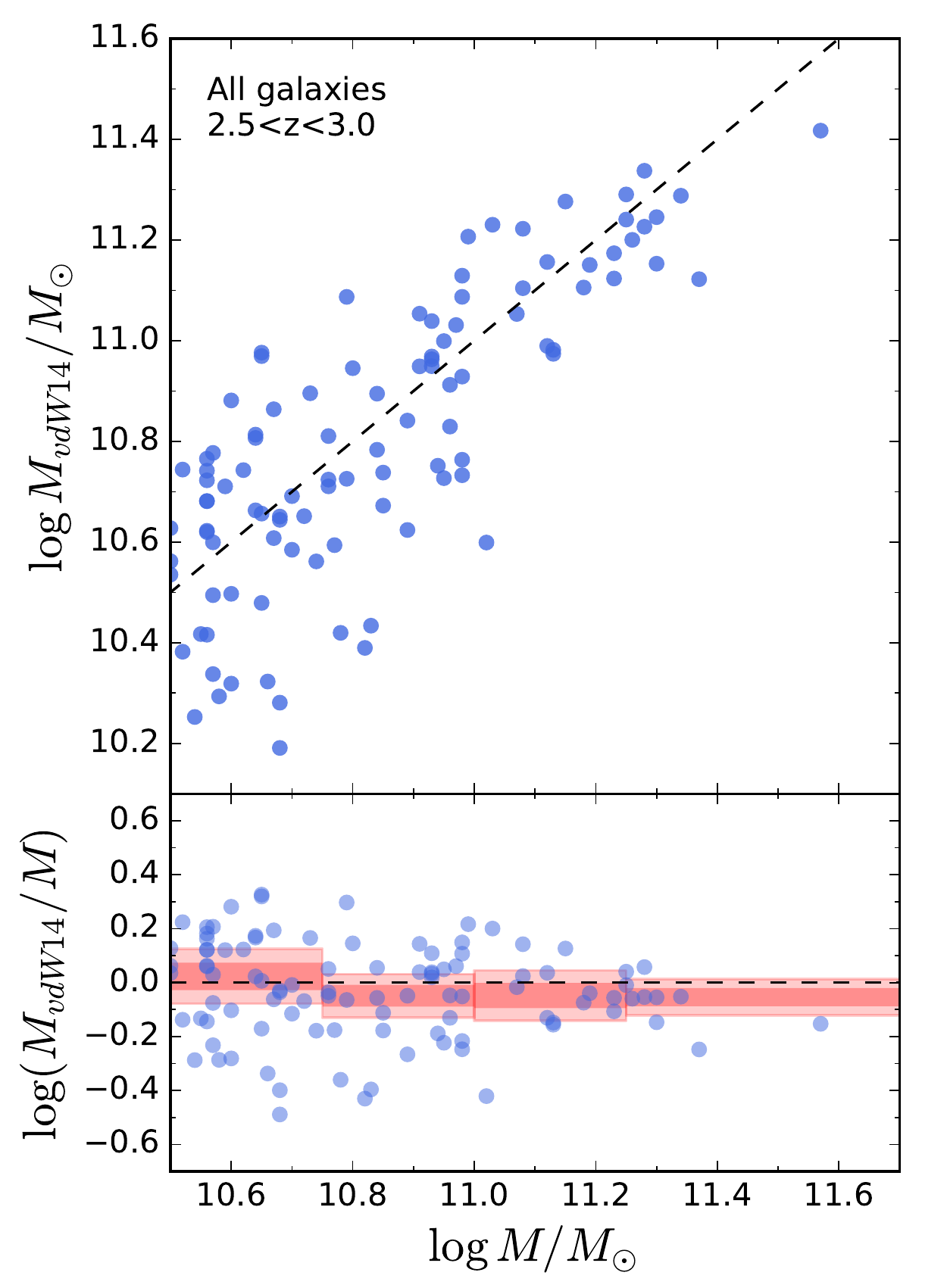}
\caption{Comparison of stellar masses used in this work from the \citet{williams2009} UDS catalog and the stellar masses used by \vdw\ from the 3D-HST UDS catalog \citep[$M_{3D}$,][]{skelton2014}.  All galaxies (i.e., quiescent and star forming) at \zwindow\  with $M>10^{10.5}$~\msun\ are shown. In the bottom panel the $1$ and $2\sigma$ limits for the median offset between the two samples are shown as red shaded regions for four bins in stellar mass.  The two mass estimates agree well.} \label{fig_mass_comparison}
\end{figure}

\subsection{Stellar masses}

In this paper we have extensively compared our sample of galaxies to those from \vdw.  Since selection by galaxy stellar mass is a key aspect of the comparison, here we show that systematic differences between the two data sets are minimal, therefore ensuring a fair comparison.

Both data sets used FAST \citep{kriek2009} to fit the SEDs and measure stellar masses assuming a Chabrier IMF.  Figure~\ref{fig_mass_comparison} shows a comparison between the stellar masses measured in our work from the \citet{williams2009} UKIDSS UDS catalog, and those used in \vdw\ from the 3D-HST catalog of \citet{skelton2014} in the CANDELS/UDS field.  All galaxies fainter than $K<24$ were selected in the former catalog at \zwindow\ and then matched to the latter catalog by RA \& Dec.  Both quiescent and star forming galaxies above $M>10^{10.5}$~\msun\ were included in order to boost the statistics.  The scatter is $\sim 0.19$~dex and the median offset of $-0.038$~dex is significant only at $\sim 1.6\sigma$.  At \masslimit\ the median offset is $-0.05$~dex with a similar low level of significance from zero.  We therefore do not apply this offset but note that doing so would not alter our conclusions given that no additional QGs from \vdw\ would enter our massive galaxy sample (see Figure~\ref{fig_sizemass}).  We also find that the stellar mass differences between the two data sets do not correlate strongly with any other relevant galaxy property such as color, redshift, or Sersic index.  Finally, any difference in stellar mass uncertainty between the two data sets is likely minimal given the lack of any strong trend seen in the bottom panel of Figure~\ref{fig_mass_comparison}.  We are therefore confidently comparing galaxies on similar stellar mass scales.

\subsection{Half-light radii}

We also verify that our half-light radii measurements are not significantly biased compared to \vdw.  We select QGs at \zwindow\ with stellar masses above $M>10^{10.7}$~\msun\ from our catalog that lie within the CANDELS UDS $H_{160}$ footprint and run them through our size measurement code in the same manner as the main sample presented in this paper.  These galaxies are then matched in RA \& Dec to objects in the structural properties catalog of \vdw, and their half-light radii compared in raw pixel units ($0\farcs06$ pixels).  The \vdw\ sizes are offset by a median of only $\sim -3.7\%$ with a scatter of $\sim 7\%$.  We do not apply this offset given its low significance level ($\sim 1.5\sigma$).

Based on the analysis above, we conclude that comparisons between the sample of QGs in this work and those in \vdw\ are not significantly impacted by systematic differences between the two data sets.

\section{Sky background estimate}
\begin{figure*}
\epsscale{1.2}
\plotone{\plotdir 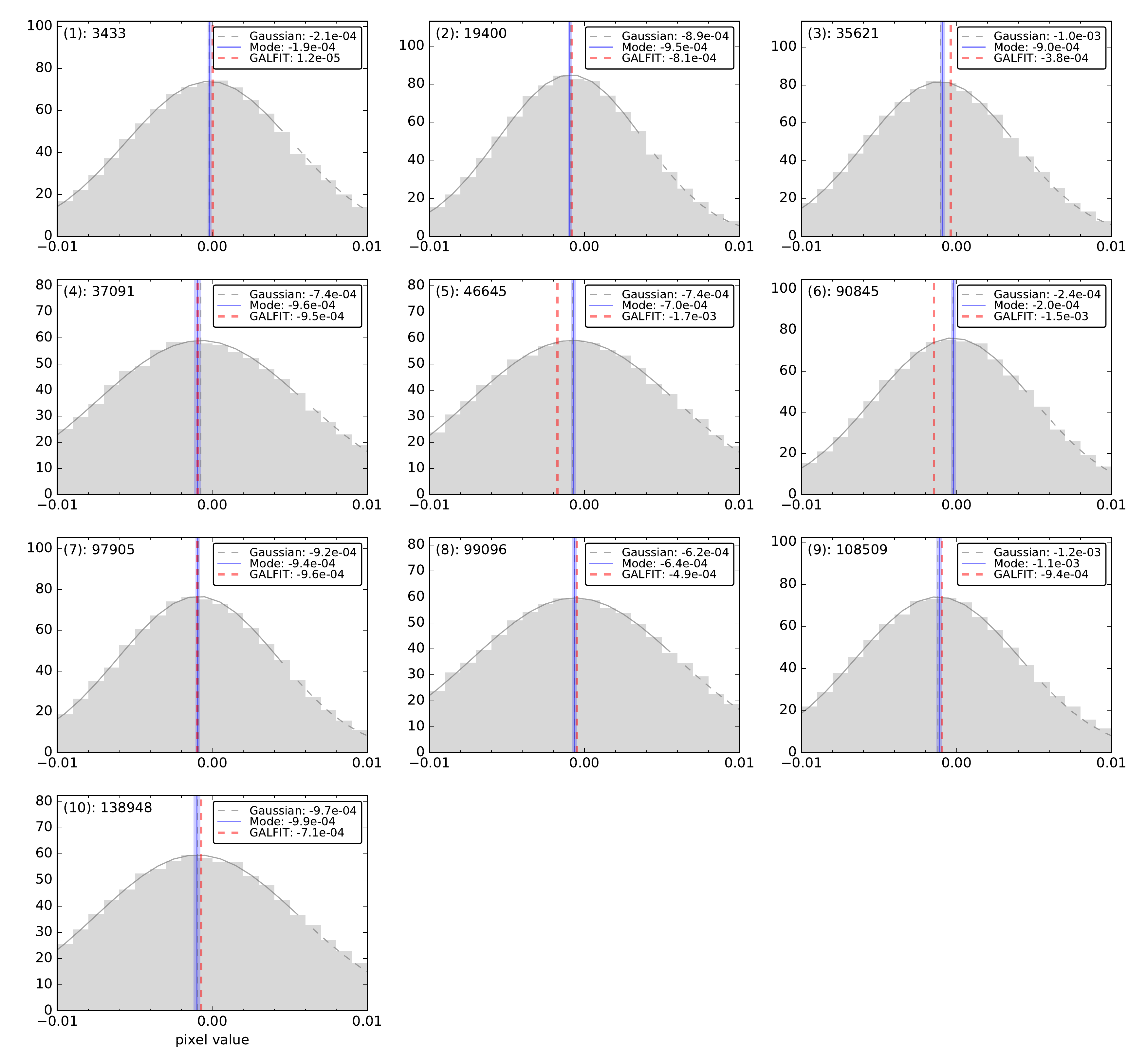}  
\caption{Estimate of the sky background level for each galaxy in our sample.  Histograms of pixel values are shown (gray) for postage stamps that are four times as big as those used in the Sersic fitting process with GALFIT.  Three estimates of the sky are shown.  They are determined from: (1) the mode of sky pixel values (blue) - our fiducial estimate, (2) fitting a Gaussian to the histogram and using its mean (gray), and (3) the GALFIT estimate based on the smaller postage stamp (red).  Method \#2 agrees well with \#1, generally falling within $\pm 2\sigma$ of the mode estimate (shaded blue region).  In the majority of cases, the GALFIT sky estimate agrees with the other estimators, but can deviate significantly in particular instances (e.g., galaxies 3, 5, 6) with the red dashed line clearly off-center from the peak of the sky distribution.  We therefore use a fixed estimate of the sky (\#1 above) when running GALFIT.} \label{fig_sky}
\end{figure*}

One of the primary sources of uncertainty in structural properties from Sersic fitting of faint high redshift galaxies is the determination of the sky background level.  The half-light radii and Sersic indices are especially sensitive to changes in the sky level \citep[e.g.,][]{bernardi2013}.  Here, we discuss our choice of employing a fixed sky level for each galaxy rather than leaving it as a free parameter for GALFIT.

The Sersic fits presented in this paper use a sky value that is the mode of sky pixel values.  We determine this mode for each galaxy by, (1) extracting a postage stamp that is twice the length (four times the area) of that used in the Sersic fitting, allowing for a more robust sky value, (2) use the HST weight map and SExtractor segmentation map to mask bad pixels and objects in this larger postage stamp, (3) sort the remaining sky pixel values, and (4) determine the pixel value that corresponds to the location where pixel values become the most finely spaced, as should be the case around the mode.  Figure~\ref{fig_sky} shows histograms of sky pixel values for each galaxy in our sample.  They are centered near zero since AstroDrizzle subtracts the sky when combining dithers.  The mode and its $\pm 2\sigma$ uncertainty are indicated by the blue line and blue shaded region.  The mode falls close to the peak of binned sky values but is not calculated from relying on an arbitrary and broad bin size.

Also shown in the figure is an estimate of the sky from fitting a Gaussian to the histogram of sky pixel values.  We include in the fit only a small portion of the positive tail (solid gray curve shows fitted region) so as to avoid any unmasked light from bright objects.  The mean of the Gaussian fit is shown by the dashed gray line.  This value agrees quite well with the mode measurement lending support to that sky value.

Finally, the red dashed line indicates the sky value when leaving it as a free parameter in GALFIT.  In most cases, it agrees reasonably well with the mode but in others it deviates significantly and is clearly off-center from the peak of the histogram of sky values.  Bright foreground galaxies likely contribute to poorly determined sky values from GALFIT, especially when the postage stamp that is used in the fitting does not extend to cover a sufficient sample of uncontaminated sky pixels.

Overall, we find that the difference in median size between using the mode vs. the GALFIT sky measurement is $<1\%$ and therefore does not impact our conclusions.  However, the scatter of $\sim 9\%$ can be driven by a small handful of objects with size differences of $\sim 30\%$ between the two methods.  The choice of which sky measurement to implement matters for such objects, and we opt to use the mode as it better represents the true sky level for all of our galaxies.




\end{document}